\documentclass[preprint]{emulateapj}
\usepackage{hyperref}
\usepackage[usenames,dvipsnames]{color}
\hypersetup{colorlinks,citecolor=Blue,linkcolor=Red,urlcolor=Blue}
\usepackage[titletoc]{appendix}
\usepackage{amsmath}

\begin{document} 

\title{An orbital window into the ancient Sun's mass}

\author{Christopher Spalding$^{1}$, Woodward W. Fischer$^2$ \& Gregory Laughlin$^1$} 
\affil{$^1$Department of Astronomy, Yale University, New Haven, CT 06511} 
\affil{$^2$Division of Geological and Planetary Sciences\\
California Institute of Technology, Pasadena, CA 91125} 

\begin{abstract}
Models of the Sun's long-term evolution suggest that its luminosity was substantially reduced 2--4 billion years ago, which is inconsistent with substantial evidence for warm and wet conditions in the geological records of both ancient Earth and Mars. Typical solutions to this so-called ``€˜faint young Sun paradox" consider changes in the atmospheric composition of Earth and Mars, and while attractive, geological verification of these ideas is generally lacking---particularly for Mars. One possible underexplored solution to the faint young Sun paradox is that the Sun has simply lost a few percent of its mass during its lifetime. If correct, this would slow, or potentially even offset the increase in luminosity expected from a constant-mass model. However, this hypothesis is challenging to test. Here, we propose a novel observational proxy of the Sun's ancient mass that may be readily measured from accumulation patterns in sedimentary rocks on Earth and Mars. We show that the orbital parameters of the Solar System planets undergo quasi-cyclic oscillations at a frequency, given by secular mode $g_2-g_5$, that scales approximately linearly with the Sun's mass. Thus by examining the cadence of sediment accumulation in ancient basins, it is possible distinguish between the cases of a constant mass Sun and a more massive ancient Sun to a precision of greater than about 1 per cent. This approach provides an avenue toward verification, or of falsification, of the massive early Sun hypothesis. 
\end{abstract}

\section{Introduction} 

The Earth has hosted life for the majority of its history \citep{Rosing1999,Schopf2007}, hinting that liquid water has persisted for billions of years. Moreover, the geological record reveals ample evidence of extensive oceans during the Archean era, 3.8--2.5\,Gya \citep{Grotzinger1993,Knoll2016}. Mars, too exhibits both ancient valley networks, carved out by running surface waters $\sim4$\,Gya \citep{Wordsworth2016}, and sedimentary basins that hosted long-lived lakes \citep{Grotzinger2015} and other large bodies of water \citep{Dibiase2013}. These features contrast sharply with Mars' modern frigid climate. Thus, from a geological perspective ancient Earth and Mars were as warm as, or warmer than today. 

\begin{figure}
\centering
\includegraphics[trim=0cm 0cm 0cm 0cm, clip=true,width=1\columnwidth]{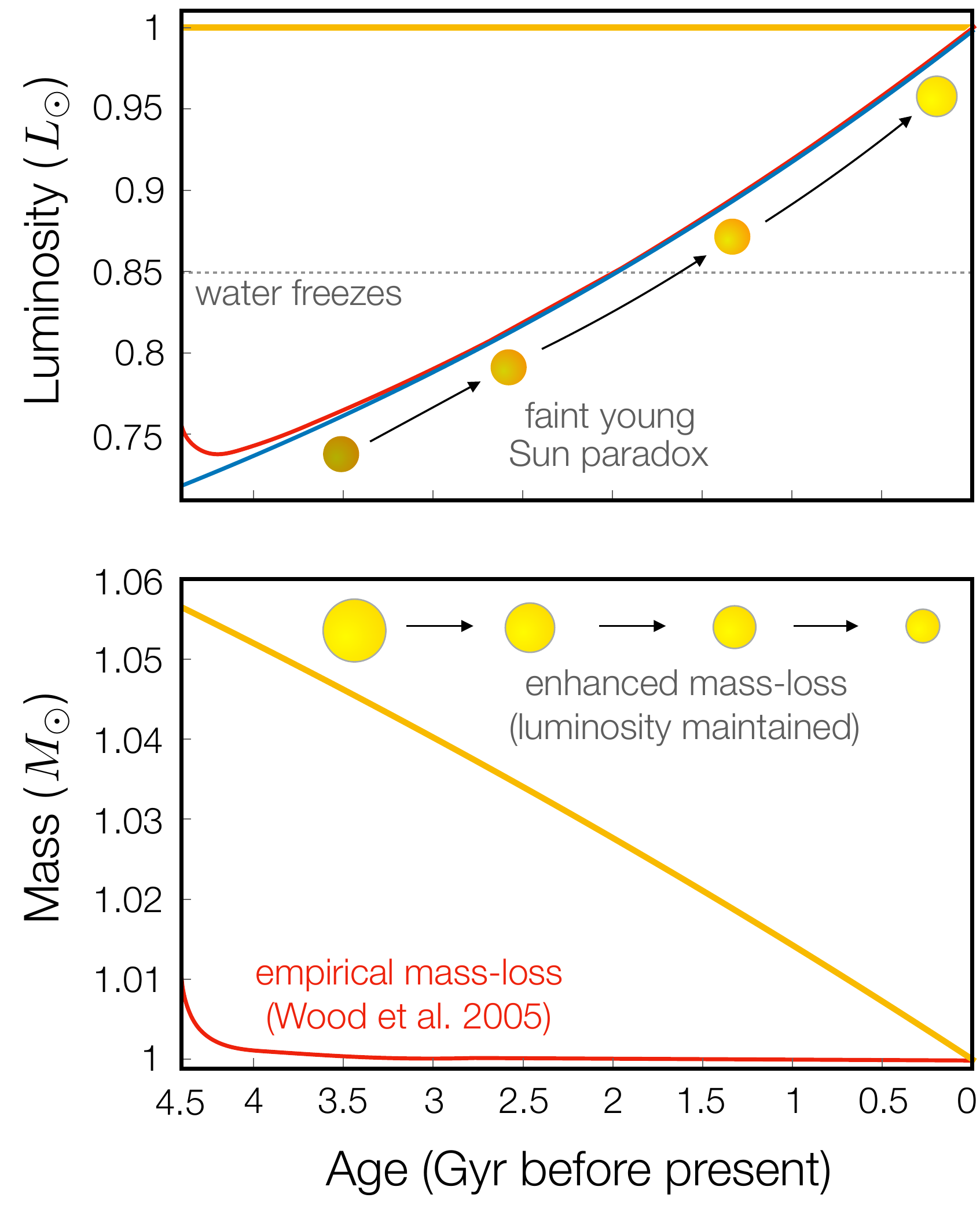}
\caption{Time evolution of the Sun's mass and luminosity under various assumptions. In the upper panel, we present the luminosity versus time as derived by \citep{Gough1981}, under the assumption of a constant stellar mass (blue line) and under a mass-loss as measured for Sun-like stars \citep{Wood2005}, in addition to a constant-luminosity model (yellow). In the bottom panel, we plot the corresponding curves illustrating the time-evolution of stellar mass, which is the property inferred from Milankovitch cycles in sediments.}
\label{Schematic}
\end{figure}

Warm climates on early Earth and Mars contrast markedly with standard models of the Sun's long-term evolution. Solar luminosity is expected to have monotonically grown throughout history, such that during the Archean era, both Earth and Mars would have received between 75--85\% as much stellar flux as today (\citealt{Gough1981}; Figure~\ref{Schematic}). All else kept equal (e.g. atmospheric composition), the Earth is expected to freeze over at stellar luminosities only $\sim10$\,\% below today \citep{Yang2012,Hoffman2017}. Given Mars' cold climate today, its ancient warmth is even more perplexing under a faint early Sun.

The apparent contradiction between the ancient Solar luminosity, and the existence of liquid water on Earth and Mars has been dubbed the ``The Faint Young Sun Paradox" \citep{Sagan1972}. Most attempts at a resolution have proposed that Earth's atmosphere contained higher concentrations of greenhouse gases \citep{Feulner2012,Charnay2013}---varying admixtures of carbon dioxide and methane, which if present in sufficient quantities may feasibly heat the Earth to a sufficient extent \citep{Bender2013}. These higher amounts of greenhouse gases would then be maintained within tight temperature bounds by the silicate-weathering feedback \citep{Walker1981}, with the supposed efficiency of this mechanism being extended to compute habitable zones around other stars \citep{Kasting1993}. The loss of habitability on Mars is thereby ascribed to the loss of its atmosphere, and therefore an efficient warming mechanism, over time \citep{Jakosky2018}.

Models incorporating high levels of greenhouse gases have attained moderate success in reproducing a warm early Earth, but it remains challenging to precisely constrain the Archean atmospheric composition using geological data. Thus, the otherwise attractive and popular hypothesis that greenhouse gases served as a solution to the Faint Young Sun Paradox, is also at present largely unfalsifiable. The extension of the problem to Mars stretches the gap between data and theory in a fashion that places serious challenges upon greenhouse gas solutions for early Martian climate. Accordingly, it is valuable to consider other hypotheses, even if only to rule them out as alternatives. 

Here, we consider the hypothesis that the evolution of luminosity derived from the standard Solar model is incorrect, because it relies on the assumption of constant mass with time. If the young Sun was a few percent more massive than at present, the stellar flux received by Earth and Mars may be maintained at similar levels to today \citep{Bowen1986,Feulner2012}. This idea is decades old, but a definitive, empirical logic for falsification has not been identified (see \citealt{Minton2007} for earlier attempts). 

We hypothesize that if the Sun lost mass over time, it would leave a recognizable fingerprint in the characteristic orbital timescales of Earth and Mars. In this work, we demonstrate that the specific period of oscillation of Earth and Mars' eccentricities, driven by Milankovitch mode $g_2-g_5$, may directly constrain the Sun's mass through time to a precision of $\lesssim1\%$. Given that these Milankovitch parameters can be preserved in the typical length scales recorded by sedimentary rocks, we now have a way that the mass of the ancient Sun can be observed. 

\section{A massive young Sun}   

Today, the Sun is losing mass predominantly by way of the stellar wind (in addition to a smaller, yet comparable mass-equivalent of photon energy). Stellar mass-loss is accompanied by an intrinsic luminosity decrease proportional to mass raised to the fourth power \citep{Chandrasekhar1939,Phillips1995}. Concomitantly, the adiabtic invariance of orbital angular momentum causes the orbits of the planets expand in inverse proportion to the Sun's mass, denoted $M_\star$. When these two effects are combined, solar mass-loss causes the flux of solar irradiation received by each planet to scale as $F\propto M_\star^6$. Consequently, in order to entirely undo the expected 25\% drop in luminosity, the young Sun requires $\sim5\%$ more mass than today. 

A larger mass for the early Sun has been proposed as a contributor to ancillary mysteries, such as the Sun's lithium depletion \citep{Graedel1991}, helioseismic signatures \citep{Guzik2010} and rotational evolution \citep{Martens2016}. These inferences of the Sun's own history are difficult to test, but attempts to measure the mass-loss from Sun-like stars might reveal crucial information regarding the Solar evolution, under the assumption that the Sun is typical.

Measurements from a small number of Sun-like stars have revealed that such stars do indeed possess stronger winds earlier in their history \citep{Wood2005}. Furthermore, there is an apparent reduction in stellar wind magnitude in stars younger than 700\,Myr that remains poorly understood. The exact magnitudes of these winds are subject to uncertainties and assumptions made in inferring their properties. Despite these issues, mass-loss rates $\dot{M}_\star$ derived as a function of time $t$ collectively follow \citep{Wood2005}
\begin{align}\label{MassLoss}
\dot{M}_\star=\dot{M}_\odot\bigg(\frac{t}{t_0}\bigg)^{-2.3},
\end{align}
where $t_0=4.5$\,Gyr is the approximate age of the Sun and $\dot{M}_\odot=2\times10^{-14}M_\odot$yr$^{-1}$ is the current Solar wind magnitude \citep{Phillips1995}.

In Figure~\ref{Schematic} we illustrate the inferred flux $F_P$ received by Earth or Mars, relative to today $F_{P,0}$, if the Sun's mass-loss followed Equation~\ref{MassLoss} (which for the sake of illustration, we extend further back than 700\,Myr after star formation, when the observations of \citet{Wood2005} become less well-understood). More specifically, we utilized the equations of \citet{Gough1981} for the evolution of the Sun's luminosity (the term within squared parentheses below), but augmented this form by a factor of $(M_\star/M_\odot)^6$, such that
\begin{align}
F_P=F_{P,0}\bigg(\frac{M_\star(t)}{M_\odot}\bigg)^6\bigg[1+\frac{2}{5}\bigg(1-\frac{t}{t_0}\bigg)\bigg]^{-1}.
\end{align}
The mass-loss prescription inferred from \citet{Wood2005} is insufficient for mass-loss of more than $\sim 1\%$, and is far short of the $\sim5$\% required to entirely remove the Faint Young Sun problem (Figure~\ref{Schematic}). Furthermore, most of the mass-loss occurs early, which may be applicable to early events in the Martian record, but does little to warm Earth's lengthy Archean era, lasting until 2.5\,Gya. 

The massive young Sun hypothesis appears at odds with observations of Sun-like stars. Nevertheless, measurements of winds from other stars are not without substantial degrees of uncertainty \citep{Vidotto2011}. Moreover, the striking dissimilarity between our Solar System and most planetary systems hosted by other stars \citep{Batalha2013} bolsters the case for our Sun experiencing an abnormal history of mass-loss. Indeed, an enhanced mass-loss may contribute to the marked absence of material interior to Mercury's orbit \citep{Hayashi1981,Chiang2013}. A more general motivation lies in the rarity with which fundamental properties of our Solar System are observable throughout antiquity\footnote{A notable example was the use of the Proterozoic-age natural nuclear reactor at Oklo, Gabon---used to constrain the ancient atomic fine structure constant \citep{Damour1996}.}. If Milankovitch records reveal the ancient Sun's mass, it is important to leverage that information.

\section{Ancient Milankovitch frequencies: Analytic expectations}

Milankovitch cycles are defined as the quasi-periodic variations of astronomically-forced insolation, first proposed as a mechanism driving Earth's past climatic variations \citep{Milankovitch1941}. We begin with an outline of the expected dependence of Milankovitch timescales upon Solar mass, based upon analytic scaling arguments. The orbits of the 8 solar system planets generally possess low inclinations and eccentricities, and their orbital periods are far from integer ratios. Consequently, the mutual gravitational interactions between planets may be approximated using a ``secular" approach. Specifically, each orbit is represented as a massive wire, exerting torques upon every other wire in the system \citep{Murray1999}. 

The time evolution of the eccentricity and inclination of planet $i$ may be written as a linear sum of oscillatory modes $g_{i}$ \citep{Laskar2004a,Laskar2011}. In the linear case, the key feature of these modes is that their oscillatory frequencies scale proportionally to the planetary orbital frequencies $n_i$, and to the ratio of the perturbing planetary mass to the Sun's mass. Thus, the linear secular frequencies scale as
\begin{align}\label{freq}
g_{i}\propto n_i=\sqrt{\frac{GM_{\star}}{a_i^3}}\frac{1}{M_\star(t)}, 
\end{align}
assuming constant planetary masses.
 
 Making the assumption that the Sun lost mass slowly compared to secular timescales, the product $M_\star a_i$ remains constant for each planet \citep{Minton2007}, such that $a_i\propto 1/M_\star$. Substituting this relationship into expression~\ref{freq} yields the proportionality
 \begin{align}\label{scaling}
 g_{i}\propto M_\star;
 \end{align}
 the linear mode frequencies scale proportionally with the Sun's mass.

The discussion above deals only with the frequencies, not amplitudes, of modes -- amplitude estimates require numerical simulations. Furthermore, we emphasize that the solar system's orbital evolution is intrinsically chaotic, making it impossible to precisely predict orbital properties further back than $\mathbf{\sim54-60}$\,Mya \citep{Laskar2011b,Zeebe2017}. Despite chaotic limitations, previous work has suggested a long-term stability of the mode associated with the frequency $g_2-g_5$, corresponding to a period of roughly 405\,Kyr today \citep{Laskar2011}. Simulations 250\,Myr into the past have shown that this mode persists as a strong driver of Earth's eccentricity oscillations.

 Below, we describe the results of a numerical simulation of the Solar system backwards 4.5\,Gyr in time. In so-doing, we tested two separate aspects of Solar System history; first is whether the stability of the $g_2-g_5$ mode persists within Earth's (and Mars') orbit into more ancient times than those considered previously \citep{Laskar2011}. Second does its frequency scale linearly with stellar mass (Equation~\ref{scaling})?

\subsection{Numerical simulations}

The linear arguments presented above ignore much complexity in the solar system's long-term evolution. The expected chaotic variations of secular modes \citep{Laskar2011}, alongside non-linear and general relativistic contributions reduce the confidence in a purely linear scaling between secular frequencies and the stellar mass \citep{Bretagnon1974}. Direct numerical simulations are required to more rigorously ascertain their mutual dependence.

\begin{figure}
\centering
\includegraphics[trim=0cm 0cm 0cm 0cm, clip=true,width=1\columnwidth]{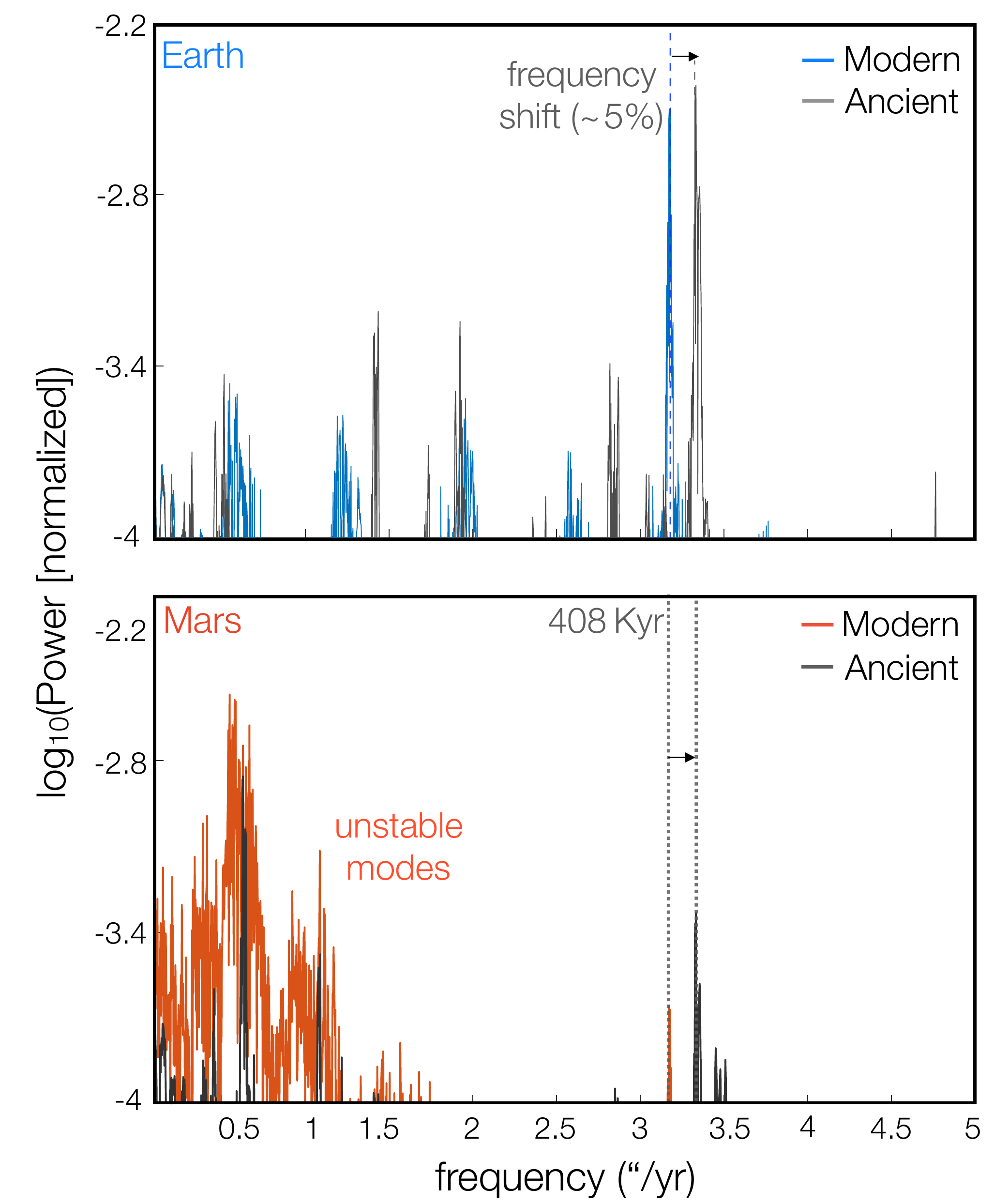}
\caption{Fourier Transform of the eccentricity evolution of Mars (bottom) and Earth (top), derived through numerical simulation. We present the most recent 450\,Myr in blue (Earth) and red (Mars), with the most ancient 450\,Myr time sequence plotted in grey (both). Notice that most modes shift somewhat, but the $g_2-g_5$ mode, labeled as ``408\,Kyr", shifts most reliably by $\sim 5\%$, i.e., by the percentage of Solar mass-loss. }
\label{f2}
\end{figure}

We performed direct \textit{N}-body simulations of the 8 solar system planets back in time 4.5\,Gyr, taking between 3 and 4 weeks to complete. A WHFast symplectic algorithm was employed within the ``\textit{Rebound}" integrator package (details in \citealt{Rein2012}). General relativistic apsidal precession was included following \citet{Nobili1986}, by way of an additional scalar potential. This prescription is simpler than that taken by \citet{Laskar2011} and leads to a small  $(\sim0.5\%$) over-estimation of secular frequencies $g_1-g_4$ (see Table~\ref{Table}) and therefore of $g_2-g_5$. Nevertheless, our primary goal is to identify the scaling between Solar mass and $g_2-g_5$, together with the mode's stability. For that, our prescription suffices. 

The Sun's mass-loss rate was set at 0.05\,$M_\odot$ per 4.5\,Gyr, such that the early Sun possessed 5\% more mass than today. As argued above, the true historic mass-loss rate was unlikely to be constant, but as long as the mass-loss timescale is long compared to secular oscillations, the exact form of mass-loss is unimportant. As initial conditions, we extracted the current orbital position for each planet from NASA's Horizons database\footnote{https://ssd.jpl.nasa.gov/horizons.cgi}. The Earth and Moon were treated together as a single object at the pair's mutual barycenter. Finally, the simulation results were tabulated every 450 years in Jacobi coordinates. The full 4.5\,Gyr output was split into 450 million-year sections, within which, the dominant frequencies were computed by way of a Fast Fourier Transform (Figures~(\ref{f2},\ref{405Kyr_Compare}) implemented in \textit{Matlab}). 

\section{Stability of the $g_2-g_5$ mode}

\begin{figure*}
\centering
\includegraphics[trim=0cm 0cm 0cm 0cm, clip=true,width=1\textwidth]{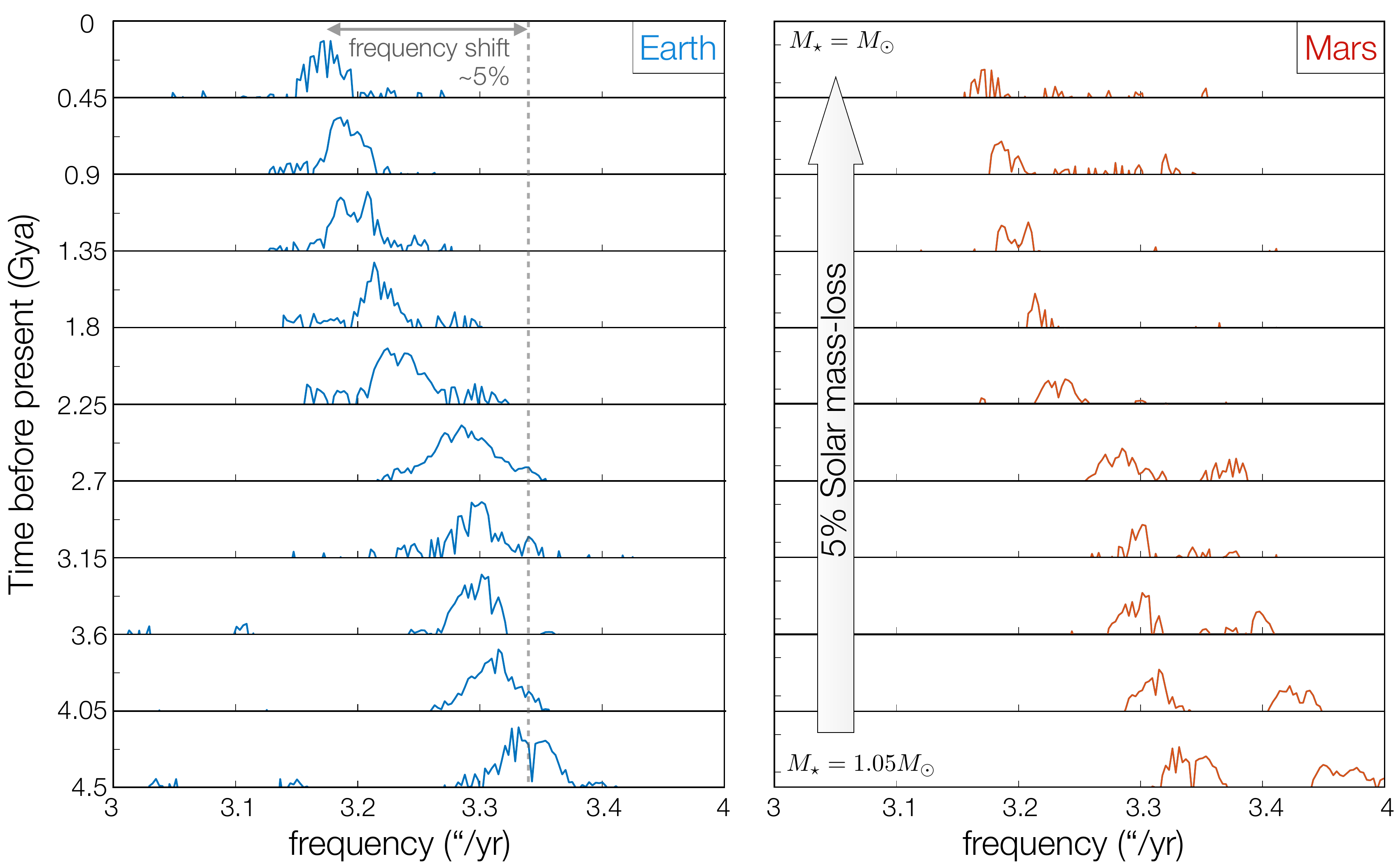}
\caption{Fast Fourier transform spectrum of the eccentricity evolution of Earth (left, blue) and Mars (right, red), within ten 450\,Myr time-bins back in time. Within each panel. the $y$-axis is an arbitrary, logarithmic scale indicating the spectral power. Time runs from bottom to top, with the Sun's mass decreasing linearly from $1.05\,M_\odot$ at the bottom to $1\,M_\odot$ at the top.}
\label{405Kyr_Compare}
\end{figure*}

Previous work has highlighted Earth's $g_2-g_5$ eccentricity cycle as the most stable of the Milankovitch frequencies \citep{Laskar2011,Hinnov2018b}, meaning that it chaotically drifts least over time. These previous results demonstrated the cycle's stability during the Phanerozoic interval, but our simulations demonstrated its stability back to 4.5\,Gya (Figure~\ref{405Kyr_Compare}). We tested whether the $g_2-g_5$ mode frequency shifts in proportion to the Sun's mass; if so, one could envision a measurement of the Sun's ancient mass. 

Figure~\ref{405Kyr_Compare} illustrates a Fourier transform of the eccentricity evolution of Earth (top) and Mars (bottom). In each plot, we present the spectrum associated with the most recent 450\,Myr section of the simulations (blue for Earth, red for Mars), alongside that for the most ancient 450\,Myr section (grey in both). The peak associated with the $g_2-g_5$ mode is highlighted; the mode shifts by $\sim5\%$ -- in proportion to the Sun's mass.

\begin{figure}
\centering
\includegraphics[trim=0cm 0cm 0cm 0cm, clip=true,width=1\columnwidth]{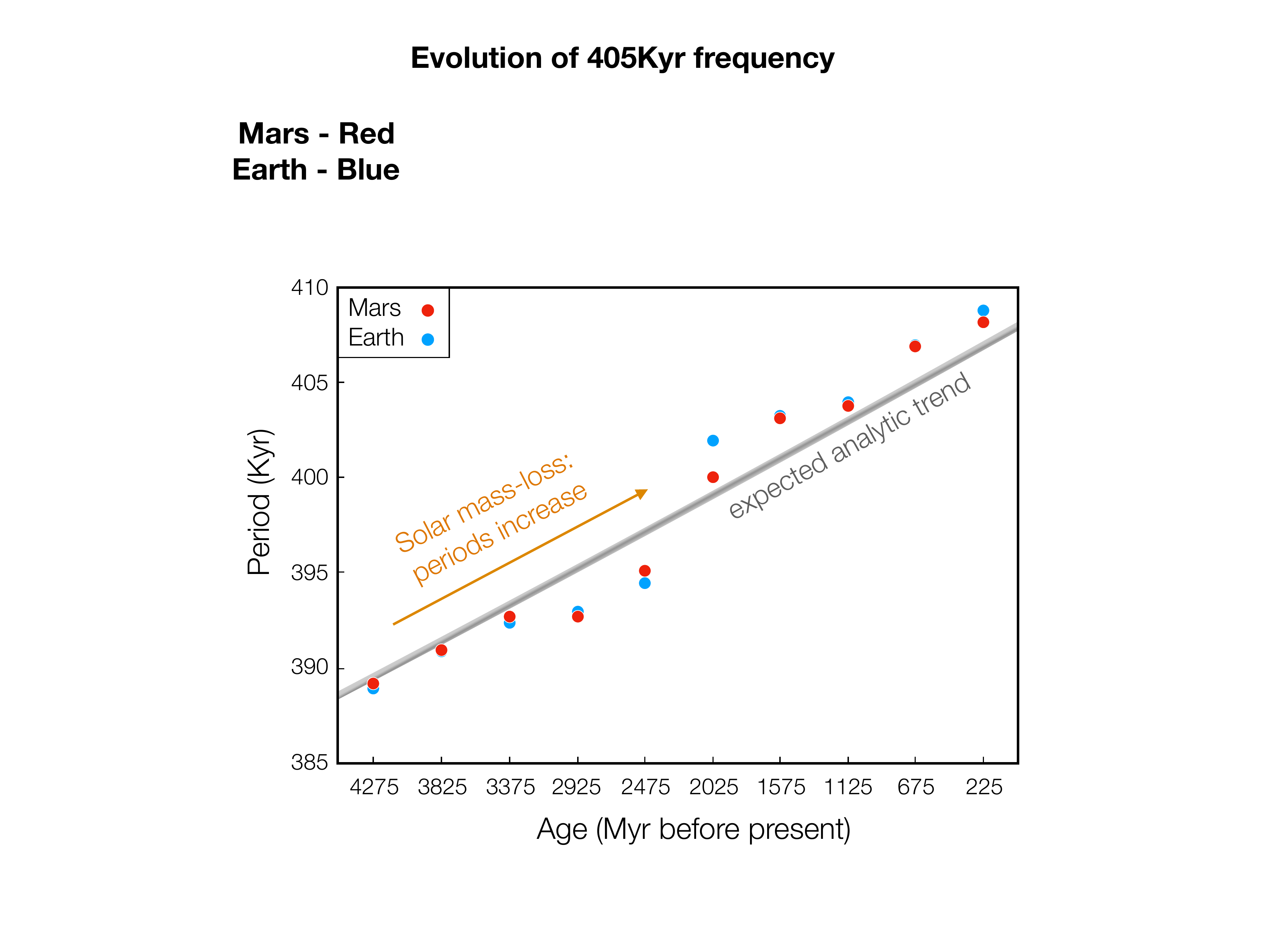}
\caption{Time evolution of the period associated with the $g_2-g_5$ mode for Mars (red) and Earth (blue). The trend closely approximates that expected from a linear proportionality between solar mass and frequency, following Equation~\ref{Relationship} (grey line). }
\label{405Kyr}
\end{figure}

Given the aforementioned chaotic nature of the Solar system, the 5\% shift of the $g_2-g_5$ mode frequency may have occurred by chance. To account for this possibility, we plotted the time-evolution of the $g_2-g_5$ mode within each successive 450\,Myr time-period (Figure~\ref{405Kyr}). Each bin is substantially longer than the $\sim60\,Myr$ divergence time, over which secular modes are predictable in detail \citep{Laskar2011b,Zeebe2017}, and so should not exhibit a predictable trend from bin to bin if the motion is dominated by chaotic drift. If, however, the time evolution is dominated by the Sun's changing mass, the period would follow a linear decrease back in time.

In Figure~\ref{405Kyr_Compare}, the peak clearly shifts to higher frequencies at earlier times. The oscillation periods corresponding to the centers of these peaks are shown in Figure~\ref{405Kyr}, with the linear relationship
\begin{align}\label{Relationship}
\boxed{|g_2-g_5|^{-1}=408\,\textrm{Kyr}\bigg(\frac{M_\star}{M_\odot}\bigg)^{-1}}
\end{align}
superimposed. Equation~\ref{Relationship} is the key result of this letter, and agrees with the simulations to $\sim1\%$. Thus, measuring the period of the $g_2-g_5$ mode in sediments of Earth and/or Mars reveals the Sun's mass to $\sim1\%$. 

Note that our simulations yield a modern period of 408\,Kyr, whereas previous work suggested 405\,Kyr. As mentioned above, this discrepancy is likely a result of our simplified treatment of relativity. In line with this explanation, we tabulated secular modes $g_1-g_8$ (Table~\ref{Table}). The modes associated with $g_1-g_4$ differ from \citet{Laskar2011} by a small amount, dominate the closer-in terrestrial planets, which are more susceptible to relativistic effects. In contrast, $g_5-g_8$ agree well with these earlier results, likely reflecting the smaller effect of relativity upon the outer planets.

Mars' orbit possesses a larger eccentricity than Earth, and is physically closer to Jupiter. As such, the degree of chaos experienced by the Martian orbit is amplified relative to Earth \citep{Laskar1994,Laskar2004b}, as can be seen by the broad range of poorly-defined peaks over its most recent 450\,Myr of evolution (Figure~\ref{405Kyr_Compare}). Thus, whereas both the orbits of Mars and Earth exhibit a predictable trend in the frequency associated with $g_2-g_5$, the amplitude of this mode is less predictable for Mars. In summary, the period of the $g_2-g_5$ signal on Mars and Earth is predictable, but the probability and strength of its occurrence on Mars is not.

\begin{table}[t]
  \begin{center}
    \begin{tabular}{| l | l | l | }
    \hline
    Mode & Value (``/yr) & La2010 (``/yr) \\ 
    \hline
 $g_1$ & 5.71&5.60  \\
$g_2$ & 7.44&7.46 \\
 $g_3$ & 17.19&17.36  \\
 $g_4$ & 17.76&17.92  \\
 \hline
 $g_5$ & 4.26&4.26  \\
 $g_6$ & 28.25&28.25  \\
 $g_7$ & 3.09&3.09  \\
 $g_8$ & 0.67&0.67  \\
\hline
    \end{tabular}
    \caption{The fundamental secular mode frequencies, computed here (middle column), compared to \citealt{Laskar2011} (La2010). Modes $g_5-g_8$ agree well. $g_1-g_4$ differ by $\sim1$\,\%, likely owing to our simplified prescription for general relativity.}
    \label{Table}
\end{center}
\end{table}

\section{Measuring ancient Milankovitch frequencies}

A link between the Earth's orbital eccentricity, climate and the banding within sediments laid down in depocenters has been suspected for centuries \citep{Imbrie1986}. These signals have now been confirmed in multiple, 10's of millions of years-long paleoclimate records, spanning most of the past 250\,Myr, and a large fraction of the earlier Phanerozoic (reviewed in \citealt{Hinnov2018b}). However, the massive-young Sun hypothesis can only be directly tested with significantly older records, extending into the early Precambrian, when high-quality strata become significantly rarer.

Some of the most ancient sources of suspected Milankovitch signals include shallow water carbonate platforms \citep{Grotzinger1986,Hofmann2004}, and banded iron formations, dating back to $\sim$2.5\,Gya \citep{Trendall2004}. In particular, Milankovitch-forced signals associated with the Earth's obliquity and precession were identified within the 2.45\,Gya Weeli Wolli Formation, and were used to constrain the moon-Earth separation.

More recent statistical techniques are emerging to better constrain frequencies recorded in ancient sediments, including that associated to $g_2-g_5$ \citep{Meyers2015,Meyers2018}. A 1.4 billion year-old sequence from Xiamaling China was used in such a way to constrain the spin rate of the early Earth. As dating methods and statistical techniques improve, a falsification of the massive early Sun hypothesis during the Archean may soon be in reach, using Earth's sediments alone.

\subsection{Sedimentary evidence from Mars}

Most of the Earth's ancient crust has been destroyed by plate tectonics and/or weathering. In contrast, Mars' relatively pristine, yet ancient sedimentary deposits offer a more transparent view of the Solar system's distant past. Roughly 40\% of the Martian surface is older than $\sim3.7$\,Gyr, with well over half exceeding $\sim3$\,Gyr in age \citep{Solomon2005}. Accordingly, most of the surface was laid down at a time during which the Sun would be expected to generate $75-80\%$ of its current luminosity at constant mass. As such, Mars may offer the most promising window in the ancient Sun's mass.

Suspected Milankovitch-forced banding of sediment has already been identified on Mars, but from orbit \citep{Laskar2004b,Lewis2008}. If future landers were equipped with precise dating techniques, most surfaces in Mars would be old enough to stand as a probe into the Sun's pasrt. A caveat, mentioned above, is that while the timescale of the $g_2-g_5$ mode reliably tracks the Sun's mass, its magnitude is less predictable. Accordingly, we cannot guarantee that any given epoch will exhibit a strong $g_2-g_5$ signal, however, if one appears, its frequency scales linearly with the Sun's mass.

Aside from Milankovitch banding, a more speculative pathway toward testing the massive early Sun hypothesis is to measure the length of a year on Mars. Orbital period scales with the inverse square of Solar mass. Daily modulations in sedimentation -- potentially from temperature-sensitive evaporite precipitation, or tidal rhythmites -- within annual cycles would constrain the number of days per year. Given the expected constancy of the Martian day length over time, the Martian year-length would immediately follow. However, more work is needed to evaluate the feasibility of such a constraint.

\section{Conclusions}

The ``Faint Young Sun Paradox" -- a discrepancy between geological evidence of warm, wet early conditions on Mars and Earth, and astronomical models suggestive of a low luminosity of the early Sun -- remains a central problem in Solar System history \citep{Feulner2012}. Its resolution would have implications for the conditions persisting during the origin of life, and the potential for life to persist elsewhere in the universe. 

The favoured solution to the Faint Young Sun paradox is typically that early Earth and Mars possessed thick, greenhouse gas-rich atmospheres, in order to trap more heat than today \citep{Sagan1972,Kasting1993,Charnay2013,Wordsworth2016}. However, an alternative possibility remains to be conclusively ruled out -- that the Sun has lost a few percent of its mass over the previous 4.5\,Gyr. In this letter, we presented a quantitative test of this hypothesis. 

Numerical simulations confirm analytic expectations that the secular frequency $g_2-g_5$ exhibits a timescale of oscillation within Earth and Mars' eccentricities that may be approximated using equation~\ref{Relationship}; the period scales linearly with the Sun's mass. A measurement of this timescale in the sedimentary records of Earth and/or Mars would provide a direct measurement of the Sun's ancient mass. 

Put succinctly, the true history of our host star's mass over billions of years has, for the first time, entered the realm of empirical investigation. 

\begin{acknowledgments}
 C.S thanks Noah Planavksy and Konstantin Batygin for useful discussions. We are grateful to the referee for a thorough report that greatly improved the manuscript. This research is based in part upon work supported by NSF grant AST 1517936, NESSF Graduate Fellowship in Earth and Planetary Sciences and the 51 Pegasi b Heising-Simons Foundation grant (C.S).
\end{acknowledgments}

\end{document}